\documentstyle[11pt,paspconf]{article}
\def\spose#1{\hbox to 0pt{#1\hss}}
\def\lta{\mathrel{\spose{\lower 3pt\hbox{$\sim$}}
    \raise 2.0pt\hbox{$<$}}}
\def\gta{\mathrel{\spose{\lower 3pt\hbox{$\sim$}}
    \raise 2.0pt\hbox{$>$}}}
\begin{document}
\title{\vspace{-10mm}THE THIRD STROMLO SYMPOSIUM --\\ THE GALACTIC 
                     HALO: BRIGHT STARS \& DARK MATTER}
\vskip -4.0truemm
\author{Tim de Zeeuw}
\vskip -2.0truemm
\affil{Leiden Observatory}
\vskip -2.0truemm
\author{John Norris}
\vskip -1.0truemm
\affil{Research School of Astronomy \& Astrophysics, 
       The Australian National University}
\bigskip
\medskip

\noindent The Third Stromlo Symposium was held at the Australian Academy of
Science in Canberra during 17--21 August 1998. The meeting was dedicated to
Alex Rodgers, who unfortunately passed away before his 65th birthday could be
celebrated in this way.  There were 110 participants from 17 countries on
five continents. Their interests broadly divided into three areas: the
{\sl stellar} halo, the {\sl gaseous} halo, and the {\sl dark} halo of our
Galaxy. The program mixed the various topics in such a way as to maximize
interaction. Together with the high-quality posters, this ensured there was
much to learn for everyone. The program also included a public lecture on
microlensing studies of dark matter by Charles Alcock, an outing to the
Tidbinbilla Nature Reserve, a wine tasting evening in the Red Belly Black
Cafe on Mount Stromlo, hosted by the incomparable Brian Schmidt, and a dinner
in the Old Parliament House which featured a number of unplanned speeches,
and an impromptu musical performance by Tim Beers (guitar) and Steve Majewski
(didgeridoo).

\section{Luminous Halo}

\vskip -2.0truemm

The stellar halo of the Galaxy contains only a small fraction of its
total luminous mass, but the abundances and kinematics of halo stars,
globular clusters, and the dwarf spheroidal satellites contain
imprints of the formation of the entire Milky Way (Eggen, Lynden--Bell
\& Sandage 1962; Searle \& Zinn 1978; Baugh, Cole \& Frenk 1996; Wyse,
White).

\medskip 
\noindent 
{\sl Abundances and ages}. Long-term systematic programs on abundances
and kinematics of halo stars (Beers, Carney) show that very few really
metal--poor stars ([Fe/H] $<-2.0$) occur in the halo. Below
[Fe/H]$=-2.0$ the numbers decrease by a factor of 10 for every dex in
[Fe/H].  Despite much effort, there is no evidence for stars
significantly more metal poor than [Fe/H]$=-4.0$ (Beers,
Norris). Stars with [Fe/H]$\lta -3.0$ display a large range in
abundances at fixed [Fe/H]. While those of C and N may result from
internal pollution (Fujimoto), that of the heavy-neutron-capture
elements probably reflects the shot noise of individual enrichment
events by early Type II supernovae, before the Type Ia's
contribute. This suggests that the most metal-poor stars formed in an
interval $\lta$1 Gyr.

New color-magnitude diagrams, an improved metallicity calibration
(Carretta \& Gratton 1997), and further work on theoretical stellar
models show that (i) nearly all globular clusters have a remarkably
small age-spread, of about 1 Gyr or less, (ii) the ages of the halo
clusters do not correlate with [Fe/H], and (iii) there is a hint that
the clusters at the largest galactocentric radii might be slightly
younger (Piotto, Vandenberg, Sarajedini).  The debate on the absolute
ages of the globular clusters continues (Mould 1998), but they
agree with those of the oldest and most metal-poor field halo stars,
based on HIPPARCOS calibrations and the Th/Eu radioactive clock
(Norris). While the dwarf spheroidal companions (Da Costa) and the
Magellanic Clouds (Olszewski) as well as M31 (Freeman) and M33
(Sarajedini) all have experienced different formation histories, it
appears that the oldest populations invariably have the same age, and
that the first generation of stars was formed synchronously throughout
the entire Local Group around 12--14 Gyr ago.
   
\medskip
\noindent
{\sl Kinematics and substructure}.  HIPPARCOS proper motions and
available radial velocities allow analysis of the space motions of a
few hundred local halo stars. There are Galactic disk stars with
$-1.6<$ [Fe/H] $-1.0$, but there is no sign of the disk for [Fe/H]$<
-1.6$, and little evidence for clumping in velocities (Chiba).  It is
not easy to interpret the results for the HIPPARCOS sample, let alone
extrapolate them to larger distances (Moody \& Kalnajs).  Hints for
velocity clumping in high-$|z|$ samples of field halo giants can be
found in many studies in the past decade (Majewski). Some satellite
dwarf galaxies may be on the same orbit, i.e., are parts of a `ghostly
stream' (Lynden--Bell). The orbits of the globular clusters display
similar correlations (Majewski).

Tidal stripping in a spherical potential results in a coherent
structure in ${\bf r}$ {\sl and} ${\bf v}$ space. This applies at
large galactocentric radii, and gives a good description of the tidal
tails of globular clusters and the Magellanic Stream (Johnston).
Tidal disruption in the flattened inner halo leaves a signature in
velocity space, but not in configuration space. Picking out disrupted
streams is possible with good kinematic data which includes
astrometry. The properties of the progenitor can then be inferred, and
in this way one can hope to reconstruct the merging history of our own
Galaxy (Helmi \& White, Harding et al.).

There was much interest in the details of the encounter of the Large
and Small Magellanic Clouds (LMC/SMC) with the Milky Way (MW). There
were talks on observations of the gas (Putman) and stars (Majewski) of
the Magellanic Stream, on theoretical models of the encounter
(Weinberg), and on the possible involvement of the Sgr dwarf
(Zhao). The total mass of the LMC may be as large as $2\times 10^{10}
M_\odot$, in which case the LMC can induce the observed warp as well
as lopsidedness in the Milky Way. In turn, the MW tidal field puffs up
the LMC, generating a stellar halo which might contribute
significantly to the microlensing event rate (Weinberg, Olszewski).

\medskip
\noindent
{\sl Gas}. The Leiden--Dwingeloo Survey (Hartmann \& Burton 1997), its
southern extension with the radio telescope in Villa Elisa, Argentina,
and the Parkes HIPASS survey have resulted in major progress in the
understanding of the high-velocity (halo) gas.  A coherent picture is
finally emerging in which (i) some HVC's are connected to the HI in
the Galactic disk, (ii) much of it is connected to the Magellanic
Stream, with the beautiful HIPASS data of Putman et al.\ (1998) now
also showing the leading arm predicted by numerical simulations of the
encounter of the Magellanic Clouds with the Milky Way, and (iii)
steady accretion of material either from the immediate surroundings of
the Galaxy (Oort 1970), or from within the Local Group (Blitz).
Absorption line studies and metallicity measurements are---at long
last---starting to constrain the distances of individual clouds
(Wakker, van Woerden).

\section{Dark Halo}

\vskip -2.0truemm

All measurements to date of the mass of the Galaxy are consistent with
a model in which the mass distribution is essentially an isothermal
sphere with a constant circular velocity $v_c$$\sim$ 180 km/s
(Zaritsky). Out to a distance of $\sim$300 kpc this corresponds to a
mass of about $2\times 10^{12} M_\odot$, and an average mass-to-light
ratio of more than 100 in solar units. The uncertainties in the halo
mass profile remain significant, even inside the orbit of the
LMC. 

\medskip
\noindent
{\sl Microlensing}. The 20 microlensing events seen towards the LMC
indicate that point-like objects in the mass range $10^{-7} \lta
M/M_\odot \lta 10^{-2}$ can at most form a minor constituent of the
Galactic halo (Alcock, Perdereau, Stubbs).  This eliminates most
objects of substellar mass, including planets as small as the
Earth. If the LMC events are caused by massive compact halo objects
(MACHOs), they must have masses in a rather narrow range around 0.5
$M_\odot$.  The unknown binary fraction in the lens population remains
a major uncertainty (Bennett). It is possible that the MACHOs are not
in the dark halo at all. Flaring of the disk and/or the warp of the
Milky Way, or the presence of another intervening object have been
considered, but it now seems unlikely that these can provide the
entire set of observed events. However, self-lensing by the LMC, in
particular by its own stellar halo, may well be quite significant
(Weinberg, Olszewski), leaving open the possibility that all the
lenses are in the LMC itself. The next generation microlensing
experiments should settle this issue (Stubbs).

\medskip
\noindent
{\sl Audit of the Universe}. The density of luminous matter
$\Omega_{\rm lum}$ is $\sim$1/10 the baryon density $\Omega_{\rm
baryon}$, which is $\sim$1/10 the total matter density of the Universe
$\Omega_M$, which in turn is $\sim$1/3 the critical density
(Turner). Recent results on distant supernovae (Schmidt et al.\ 1998;
Perlmutter et al.\ 1999) suggest that the `dark energy'
$\Omega_\Lambda$ brings the total $\Omega_0 = \Omega_M +
\Omega_\Lambda$ to the critical value. The value of $\Omega_{\rm
baryon}$ is of the same order as $\Omega_{\rm galaxies}$, suggesting
that galactic halos consist of dark baryons, and that non-baryonic
dark matter is needed at the scales of clusters and larger in order to
provide $\Omega_M$. If only one kind of dark matter exists on all
scales, then the location of most of the baryons needs to be found.

\medskip
\noindent
{\sl Composition}.  Many talks addressed the nature of dark matter.
The measured number density of low-mass objects shows that brown
dwarfs cannot provide all the dark mass in the Galactic halo (Flynn,
Tinney). The microlensing events towards the LMC suggest white dwarfs
as major constituent of the dark halo (Chabrier). The required numbers
imply a special initial mass function early-on, and a resulting
metal-enrichment which is hard to hide (Gibson \& Mould 1997).
Neutron stars face similar problems (Silk).

Ultracold (4K) clumps of H$_2$, with masses of $10^{-3} M_\odot$,
diameters of 30 AU, and densities of $10^{10}$ cm$^{-3}$ have also
been proposed as dark matter candidates.  Current microlensing
experiments are not sensitive to such clumps, but they have not been
seen directly in the local interstellar medium. They might reside in a
large outer disk (Pfenniger), or in a spheroidal halo (Walker). The
ionized and evaporating outer envelopes of the clumps could cause
extreme scattering events at radio wavelengths (Fiedler et al.\ 1987),
but it is unclear whether the expected cosmic-ray induced gamma rays
are seen (Chary; Dixon et al.\ 1998).

Ionized gas at $10^6$ K may be the best bet for baryonic dark matter
attached to the Galaxy (Fukugita, Hogan \& Peebles 1998).  The current
observational constraints on the total amount of this material
(Maloney) are no stronger than they were 40 years ago (Kahn \& Woltjer
1959), but ROSAT (Kalberla) and ongoing H$\alpha$ surveys
(Bland--Hawthorn) will allow significant improvement soon.

Candidates for non-baryonic dark matter include massive neutrinos,
axions, neutralinos, and primordial black holes (Sadoulet, Silk,
Turner). Neutrinos are not favored by theories of structure formation,
and the required mass of $\sim$25 eV may be ruled out.  There is no
experimental evidence for axions or neutralinos, but the laboratory
sensitivity will improve considerably in the coming year.

\section{Towards a stereoscopic census} 

\vskip -2.0truemm

A quantitative test of formation scenarios for the Galaxy is a crucial
complement to high-redshift studies of galaxy formation. It will
require accurate distances and kinematic data for large samples of
halo objects. We can expect ongoing objective prism surveys
(Christlieb) to identify $\sim$20000 candidate halo objects.
Multi-object spectroscopy by 2DF or SLOAN will provide radial
velocities and abundances (Pier). The Tycho Reference Catalog will
contain proper motions of 2--3 mas/yr for 3 million stars to
$V$$\sim$12 (Hoeg et al.\ 1998), allowing space motions to be
determined for $\sim$30000 halo stars out to a few kpc.

The next major step will have to await NASA's Space Interferometry
Mission, and GAIA, a mission under study by ESA (Gilmore et al.\
1998).  SIM and GAIA will obtain proper motions and parallaxes to
$\lta$10 micro-arcsecond accuracy. While SIM will be a pointed
observatory, GAIA will measure {\it all} one billion stars to
$V$$\sim$20, and obtain radial velocities and photometry for most
objects brighter than $V$$\sim$17.  GAIA will provide the full
six-dimensional phase space information (positions and velocities) to
$\sim$20 kpc from the Sun, and velocity information on individual
stars to much larger distances.  High-resolution ground-based
spectroscopic follow-up will provide abundances throughout the halo,
so that the entire formation history of the Galaxy can be
reconstructed.

\end{document}